\documentclass[a4paper,english,12pt]{article}
\usepackage[dvips]{graphicx}
\usepackage{hyperref,amsmath,amssymb}
\usepackage[footnotesize,sc]{caption2}

\newcommand{\eq}[1]{eq.\ (\ref{#1})}

\newcommand{\fig}[1]{Fig.\ \ref{#1}}
\newcommand{\tab}[1]{Table \ref{#1}}

\newcommand{\rO}{\mathrm{O}}
\newcommand{\mathe}{\mathrm{e}}
\newcommand{\tmop}[1]{\operatorname{#1}}

\title{{\normalsize\vskip -50pt
\mbox{} \hfill HU-EP-05/26 \\
\mbox{} \hfill DESY 05-091 \\
\mbox{} \hfill SFB/CPP-05-21 \\}
\vskip 25pt
Cutoff Effects in O($N$) Nonlinear Sigma Models}
\author{
	Francesco Knechtli \raisebox{1ex}{\footnotesize a},
	Bj\"orn Leder \raisebox{1ex}{\footnotesize b} and
	Ulli Wolff \raisebox{1ex}{\footnotesize a}
}
\date{\small\it
\raisebox{1ex}{\footnotesize\rm a} Institut f\"ur Physik, Humboldt Universit\"at, Newtonstr. 15, 12489 Berlin, Germany\\
\raisebox{1ex}{\footnotesize\rm b} DESY Zeuthen, Platanenallee 6, 15738 Zeuthen, Germany
}

\begin{document}

\maketitle
\thispagestyle{empty}
\setcounter{page}{0}

\begin{abstract}
In the nonlinear O($N$) sigma model  at $N=3$
unexpected cutoff effects have been found before
with standard discretizations and lattice spacings.
Here the situation is analyzed further employing additional data
for the step scaling function of the finite volume massgap
at $N=3,4,8$ and a large $N$-study 
of the leading as well as next-to-leading terms in $1/N$.
The latter exact results are demonstrated to follow
Symanzik's form of the asymptotic cutoff dependence.
At the same time, when fuzzed with artificial statistical
errors and then fitted like the Monte Carlo results, a 
picture similar to $N=3$ emerges.
We hence cannot conclude a truly anomalous cutoff dependence
but only relatively large cutoff effects, where the logarithmic
component is important. Their size shrinks at larger $N$, but
the structure remains similar.
The large $N$ results are particularly interesting as we here
have exact nonperturbative control over an asymptotically free
model both in the continuum limit and on the lattice.

\end{abstract}

\clearpage

\section{Introduction}
The O($N$) nonlinear sigma models in two space time dimensions are
a very precisely studied class of quantum field theories in many respects.
For $N\ge 3$ they are asymptotically free like QCD and hence furnish
a simplified test laboratory for all kinds of questions. This includes the
question of the dependence of physical results on the lattice spacing and this
is the subject studied in this publication.

While we are interested in results at zero lattice spacing, nonperturbative
Monte Carlo results are always at finite $a$. One hence has to assume an analytical
form for the asymptotic $a$-dependence to extrapolate to zero. Ideally this behavior
is checked with good precision for a large range of $a$ and then assumed for the
`remaining' extrapolation. In principle a systematic error has to be estimated here,
which however is often deemed to be negligible compared with statistical errors, which
tend to be amplified in the extrapolation. This balance depends on the `extrapolation fit'
one uses. If it is restrictive with few parameters (but compatible with the data) statistical
errors remain small but the danger of systematic effects is high. In the opposite case
one may `give away' statistical precision and overestimate total errors. It is presumably clear from
these words that in the extrapolation step there is some danger of subjective elements getting
into a presumed `first principles' calculation.
The situation is aggravated in QCD with dynamical fermions, because due to algorithmic and CPU-power
limitations one cannot vary $a$ over a large enough range. Continuum results then mainly
arise under the {\em assumption} of a certain $a$-dependence.

Almost universally in lattice field theory the model of the asymptotic cutoff dependence
is derived from Symanzik's analysis of lattice perturbation theory to all orders \cite{Symanzik:Cutoff}.
It extends the investigation of renormalizability to subleading orders in the cutoff 
and is assumed to hold in structure also beyond perturbation theory and to thus ascertain the existence
of the continuum limit and predict the rate at which it is approached. The results for the
rate are powers in $a$ (usually $a^2$ or linear for fermions) modulated by only slowly varying
logarithmic factors. 

In view of this situation there have been efforts to explore this in detail in the
sigma model lab. The situation there is vastly better than in QCD: 
No critical slowing down due to
cluster algorithms \cite{Wolff:1988uh} , 
well-understood and well-measurable finite size quantities \cite{stepscaling},
reduced variance estimators \cite{Hasenbusch:1994rv}. The Symanzik prediction for this bosonic theory is
clearly $a^2$ scaling. In earlier investigations \cite{Hasenfratz:2001,Hasenbusch:2002}
it has been found that some results
of the O(3) model do not fit very well with this standard expectation down to rather small $a$-values.
These studies are extended here in the direction of larger $N$-values, since it is known analytically that
eventually at $N=\infty$ the Symanzik picture holds. We here report in Sect.~2 Monte Carlo results
for $N=3,4,8$ with the standard lattice discretization. In Sect.~3 we investigate the large $N$ expansion
including the leading correction of order $1/N$. In Sect.~4 we draw
some conclusions.

\section{Monte Carlo Data}\label{mc}

    \begin{figure}[tb]
	    \centering
	    \includegraphics{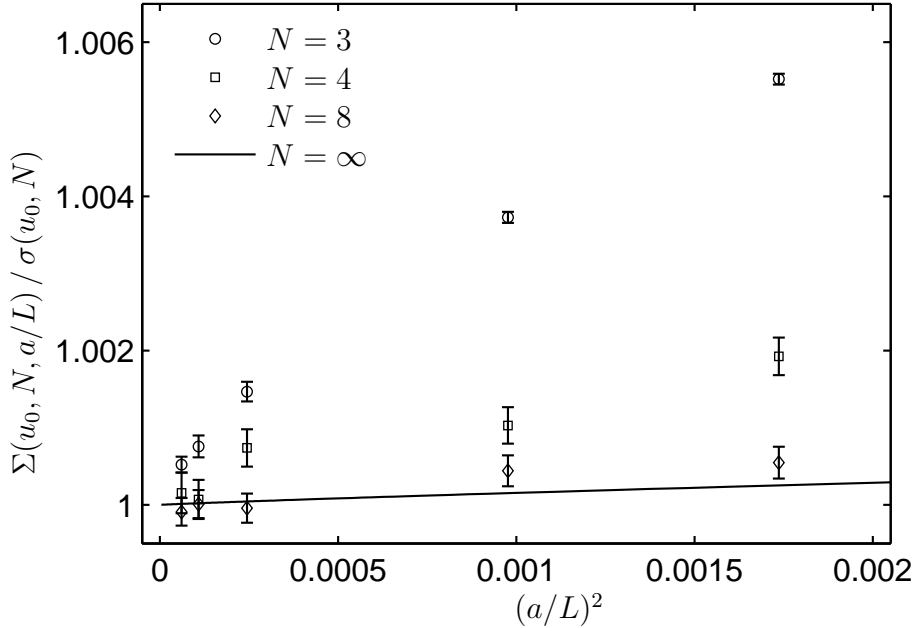}
	    \caption{Lattice step scaling function $\Sigma(u_0,N,a/L)$ normalized by the continuum value
	    		$\sigma(u_0,N)$ at $u_0=1.0595$.}
	    \label{fig_cutoff}
    \end{figure}

We consider the
lattice O($N$) nonlinear $\sigma$-model
\begin{equation}
  Z = \int \prod_x d^N s \delta ( s^2 - 1 ) \mathe^{- S ( s )}
  \label{Zsigma}
\end{equation}
with the standard lattice action
\begin{equation}
  S ( s ) = \frac{N}{2 \gamma} \sum_{x \mu} ( \partial_{\mu} s )^2
\end{equation}
where $\partial_{\mu}$ is the forward difference operator and $\gamma$ the rescaled
coupling (fixed at large $N$).
Both in our MC simulations and the large $N$ expansion we determine
the step scaling function (for a factor two rescaling)
\begin{equation}\label{Sigma}
    \Sigma(u,N,a/L) = M(2L)\,2L\,, \qquad u=M(L)\,L\,,
\end{equation}
where $M(L)$ is the mass-gap \cite {Luscher:1982uv}
of the transfer matrix. It is determined
from the asymptotic exponential fall-off of the two-point function at
zero spatial momentum. In principle this refers to a periodic box with finite spatial size
$L$ and infinite extent in the time direction. In the simulations
this was approximated as described in \cite{stepscaling}.
The renormalized coupling $M(L)L$ used here differs by a 
factor $(N-1)/2$ from $\bar{g}^2(L)$ in \cite{stepscaling}
to keep it finite in the large $N$ limit. 
The lattice step scaling function
is expected to reach a finite continuum limit,
\begin{equation}\label{sigma}
 \lim_{a/L\to0}\Sigma(u,N,a/L)=\sigma(u,N)
\end{equation}
and we discuss here how this is approached.
At large $N$ we shall use the expansion
\begin{equation}\label{largeNSigma}
 \Sigma(u,N,a/L)=\Sigma_0(u,a/L) + \frac1{N} \, \Sigma_1(u,a/L) +\rO(N^{-2}).
\end{equation}

\begin{table}[htb]
	\centering
	\begin{tabular}{rccccc}
	$L/a$	& $\Sigma(u_0,3,a/L)$	& $\Sigma(u_0,4,a/L)$ & $\Sigma(u_0,8,a/L)$ & 
                 $\Sigma_0(u_0,a/L)$ & $\Sigma_1(u_0,a/L)$ \\ \hline
	5	& 1.29379(8)	&		& -	      &  -        & -       \\ 
	6	& -		& 1.31256(47)	& 1.35065(40) &  1.379491 &-0.194123 \\ 
	8	& -		& 1.31068(43)	& 1.34891(35) &  1.378215 &-0.196201 \\
	10	& 1.27994(9)	& 1.30825(39)	& 1.34835(34) &  1.377524 &-0.197668 \\ 
	12	& 1.27668(9)	& 1.30608(37)	& 1.34721(32) &  1.377112 &-0.198665 \\ 
	16	& 1.27228(12)	& 1.30477(34)	& 1.34725(30) &  1.376665 &-0.199867 \\ 
	24	& 1.26817(9)	& 1.30379(32)	& 1.34649(28) &  1.376309 &-0.200939 \\ 
	32	& 1.26591(9)	& 1.30263(31)	& 1.34635(27) &  1.376170 &-0.201394 \\ 
	64	& 1.26306(16)	& 1.30225(32)	& 1.34570(25) &  1.376022 &-0.201921 \\ 
	96	& 1.26216(18)	& 1.30138(33)	& 1.34577(24) &  1.375990 &-0.202042 \\ 
	128	& 1.26187(13)	& 1.30149(34)	& 1.34563(24) &  1.375978 &-0.202089 \\ \hline 
	cont.	& 1.261208(1)	& 1.3012876(1)	& 1.345757(2) &  1.375961 &-0.202161 \\ \hline 
	\end{tabular}                                          
	\caption
		{Monte Carlo and large $N$ data for the lattice step scaling
		function at $u_0=1.0595$. }
	\label{datatab}
\end{table}

For technical details about our algorithm and estimators used in the simulations we refer the reader
 to Ref. \cite{Leder:2003}. Results are listed in \tab{datatab}. The O(3) data for $L/a\le64$ has been 
taken from
\cite{Hasenbusch:2002}, where  Seefeld et. al. study 
cutoff effects in $\Sigma(u_0, 3, a/L)$ at the popular value
$u_0=1.0595$ already appearing in \cite{stepscaling}.
We have extended their data
to the lattices $L/a=96,\,128$ and have added results at $N=4,8$ to explore the transition
to the large $N$ behaviour that we later study semi-analytically.
The bottom line of the table contains proposed exact continuum values
published in \cite{Balog:2003yr} for $N=3$ and the values for $N=4,8$ \cite{BalogPC}. 
They depend on some mild theoretical assumptions, but look
consistent with our numerical values \cite{Leder:2003,Balog:2003yr}.
More precisely, an extrapolation with the fit-forms considered below 
(eqns. (\ref{fitaopt}) 
-- (\ref{fitlog2}))
leads to extrapolated values that agree with the conjectured exact
values within errors (propagated from the Monte Carlo statistical errors)
as soon as the $\chi^2$ of the fit is acceptable\footnote{
There is one exception to this for $N=3$: 
a free exponent fit (form (\ref{fitaopt})) including 
coarser lattices,
where the conjectured value is missed by several standard deviations with an exponent close to one.}.
More details can be found in \cite{Leder:2003}.
In this publication {\em we assume the conjectured continuum
values to be correct} and use them
to constrain our extrapolations. We notice that the case $N=8$ is very well described
by the large $N$ expansion.
Values of $\Sigma$ at $N=8$ and $N=\infty$
differ by about 2.2\%. This discrepancy is lowered to about $0.4\%$ by the $1/N$
correction that we shall derive in this paper.

We have a first look at cutoff effects in \fig{fig_cutoff} where we simply
plot $\Sigma/\sigma$ versus $(a/L)^2$
for the different values of $N$ together with the exact curve for $N=\infty$.
Again $N=8$ is already close to the large $N$ limit.
The lattice artifacts for 
$N>3$ generally turn out to be much smaller than for $N=3$,
which makes the numerical investigation of their structure
more difficult. While the $N=3$ data look curved to the naked
eye in this representation, this cannot be decided so easily
for the higher $N$-values.
In order to analyze the cutoff effects in more detail
we will fit several analytic forms to the data. In the following Section we
explain what kind of fits we have done and how and what we can learn from them.

\subsection{Fits to the data}

Symanzik's analysis of the cutoff dependence of lattice Feynman diagrams
suggests that leading lattice artifacts are quadratic in the lattice spacing modified by slowly varying
logarithmic functions. It should be noted however that logarithms generically appear in an $l$'th
order polynomial at $l$ loop order. If the series were summable the leading power could thus
be modified. Nevertheless in typical lattice applications it is assumed that the integer
power behaviour is modified only weakly and that this can be neglected for most attainable
statistical precisions in an extrapolation.
It was first found in \cite{Hasenfratz:2001} that in the O(3) model
instead of the expected quadratic dependence a behaviour more close to a linear
is suggested. With this background we first investigate fits with a general exponent
$\alpha$
\begin{equation}\label{fitaopt}
    \mathrm{A}: \quad  c_0 + c_1\, (a/L)^{\alpha} \,,
\end{equation}
where $c_0$ is fixed by our exact continuum data. Such a fit results in an optimal exponent
$\alpha_{\mathrm{opt}}$ that minimizes the $\chi^2$ deviation between (\ref{fitaopt}) and the data. 
To judge the plausibility of such a fit we compute the probability $Q$
of finding a value of $\chi^2$ larger or equal to the one encountered in the fit at hand
(goodness of fit test \cite{Numrec}). In general, values $Q<0.1$ are considered implausible.

    \begin{figure}[tb]
	    \centering
	    \includegraphics{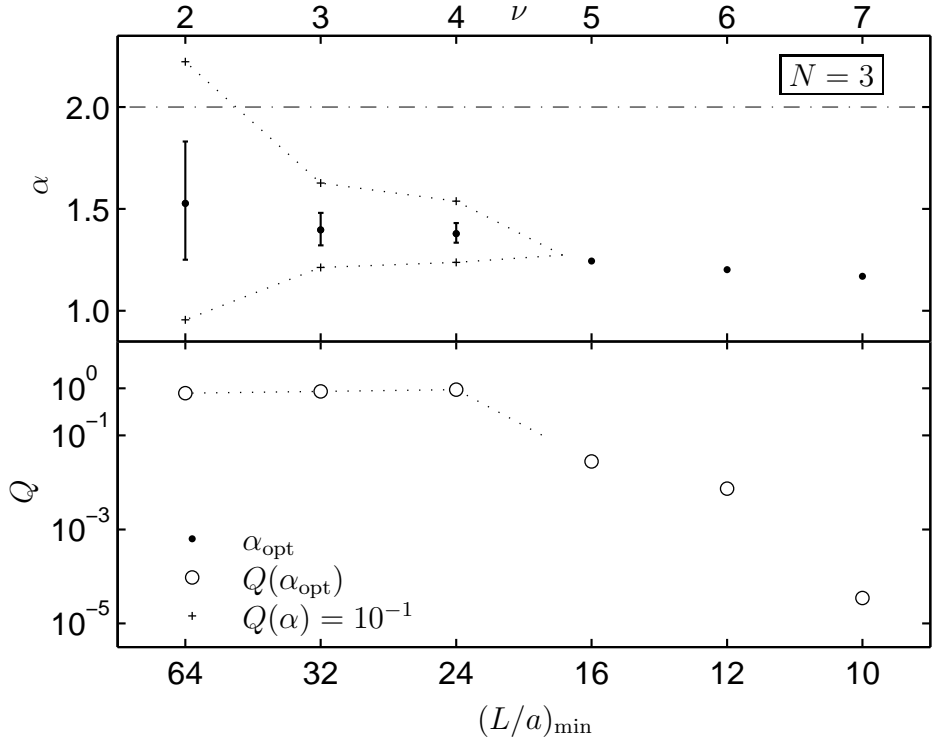}
	    \caption{Results of the fit to \eq{fitaopt} for $N=3$. In the upper half the optimal 
	    exponent and the limits where the fit with fixed $\alpha$ becomes unlikely is plotted.
	    The lower half shows $Q(\alpha_{\mathrm{opt}})$. The dotted lines are to guide the eye.}
	    \label{fig_ssf3_aopt_imp}
    \end{figure}

    \begin{figure}[tb]
	    \centering
	    \includegraphics{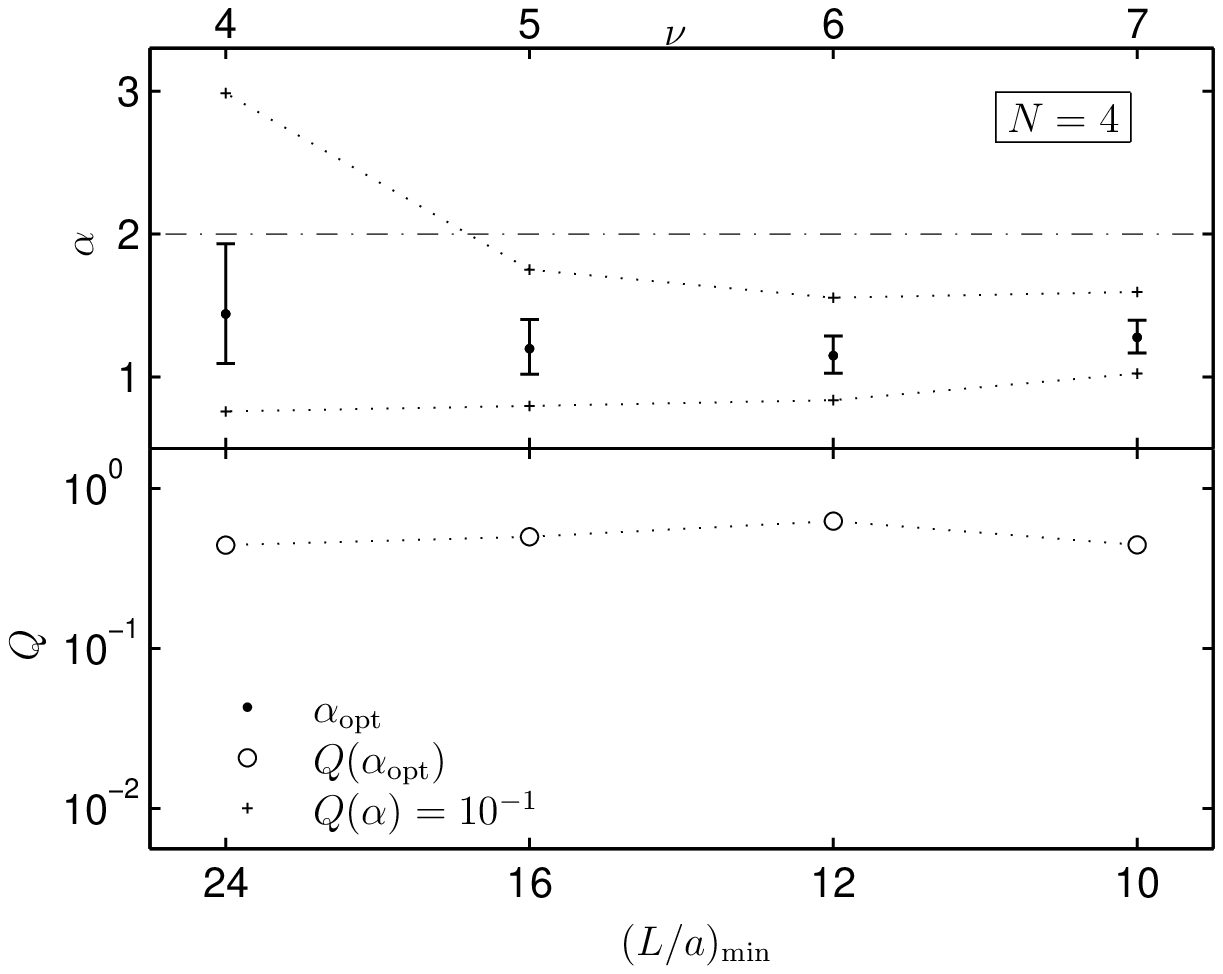}
	    \caption{As \fig{fig_ssf3_aopt_imp} but for $N=4$.}
	    \label{fig_ssf4_aopt_imp}
    \end{figure}

    \begin{figure}[tb]
	    \centering
	    \includegraphics{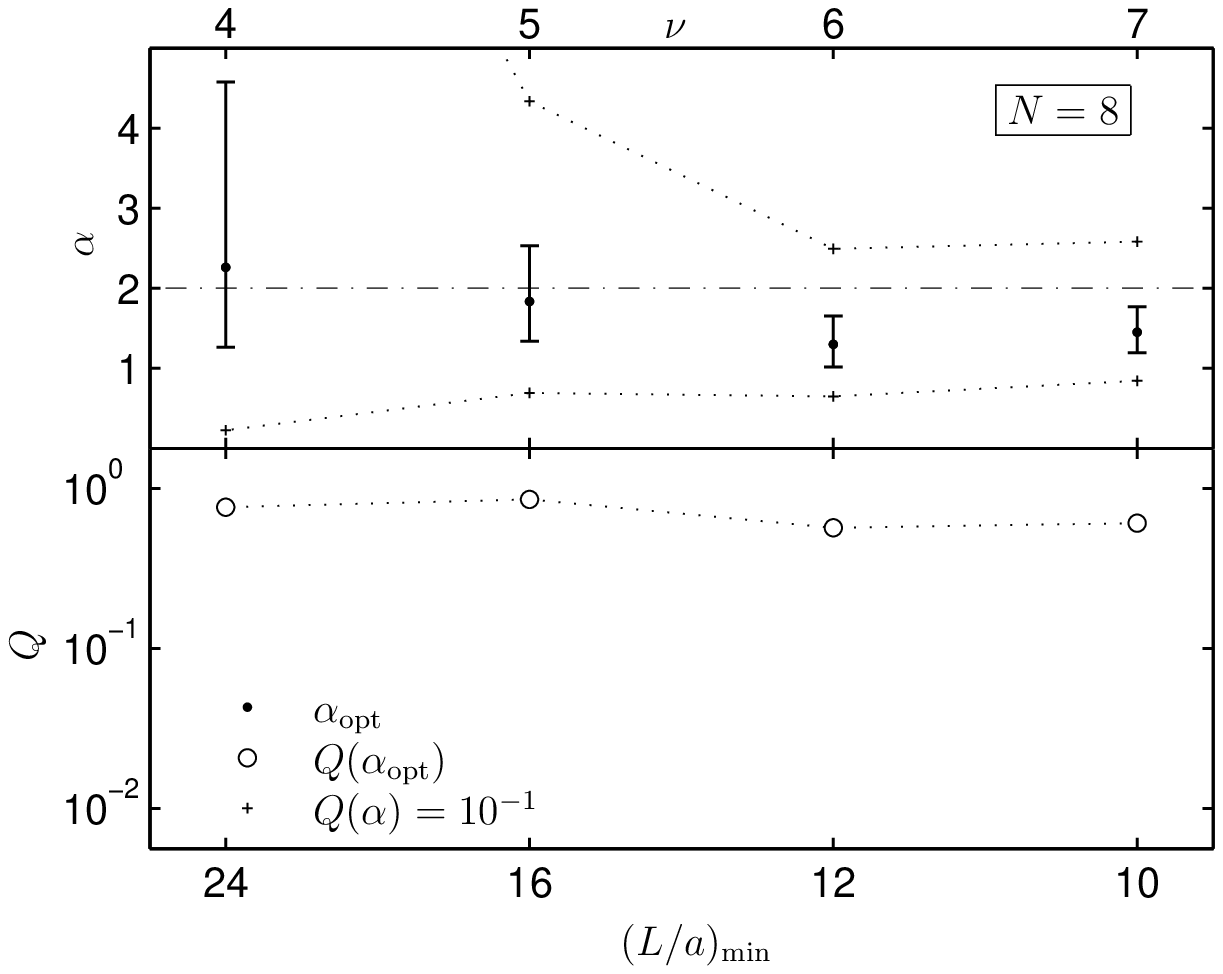}
	    \caption{As \fig{fig_ssf3_aopt_imp} but for $N=8$.}
	    \label{fig_ssf8_aopt_imp}
    \end{figure}

For $N=3$ fit results are shown in \fig{fig_ssf3_aopt_imp} which requires some explanation.
Since all theoretical forms of 
the cutoff behavior are only
asymptotic, we successively exclude coarser lattices from the fit. We therefore find values
$\alpha$ and $Q$ as functions of $(L/a)_{\mathrm{\min}}$, the size in lattice units
of the coarsest lattice still included
in the fit.
The top labels list the numbers $\nu$ of the
remaining degrees of freedom in the fit.
Errors of $\alpha_{\mathrm{opt}}$ are assessed in two ways. The errorbars are limited by those 
values of $\alpha$
where the optimal $\chi^2$ grows by unity. The plus signs are the limits where $Q$ falls below 0.1.
When there are no fits with $Q\ge0.1$ no errors are drawn.
We see here that acceptable fits are only achieved with lattices $L/a \ge 24$.
Moreover they are neither consistent with the Symanzik value $\alpha=2$ (dash-dotted line)  nor
with $\alpha=1$ but rather favor values in between. For $N=4,8$ the equivalent plots are
shown in \fig{fig_ssf4_aopt_imp} and \fig{fig_ssf8_aopt_imp}. In these cases the proposed
fit is acceptable for all $(L/a)_{\mathrm{\min}} \ge 10$. For $N=4$ the preferred $\alpha$-values
are still smaller than two, but $N=8$ starts to look consistent with the whole range
between one and two.

    \begin{figure}[tb]
	    \centering
	    \includegraphics{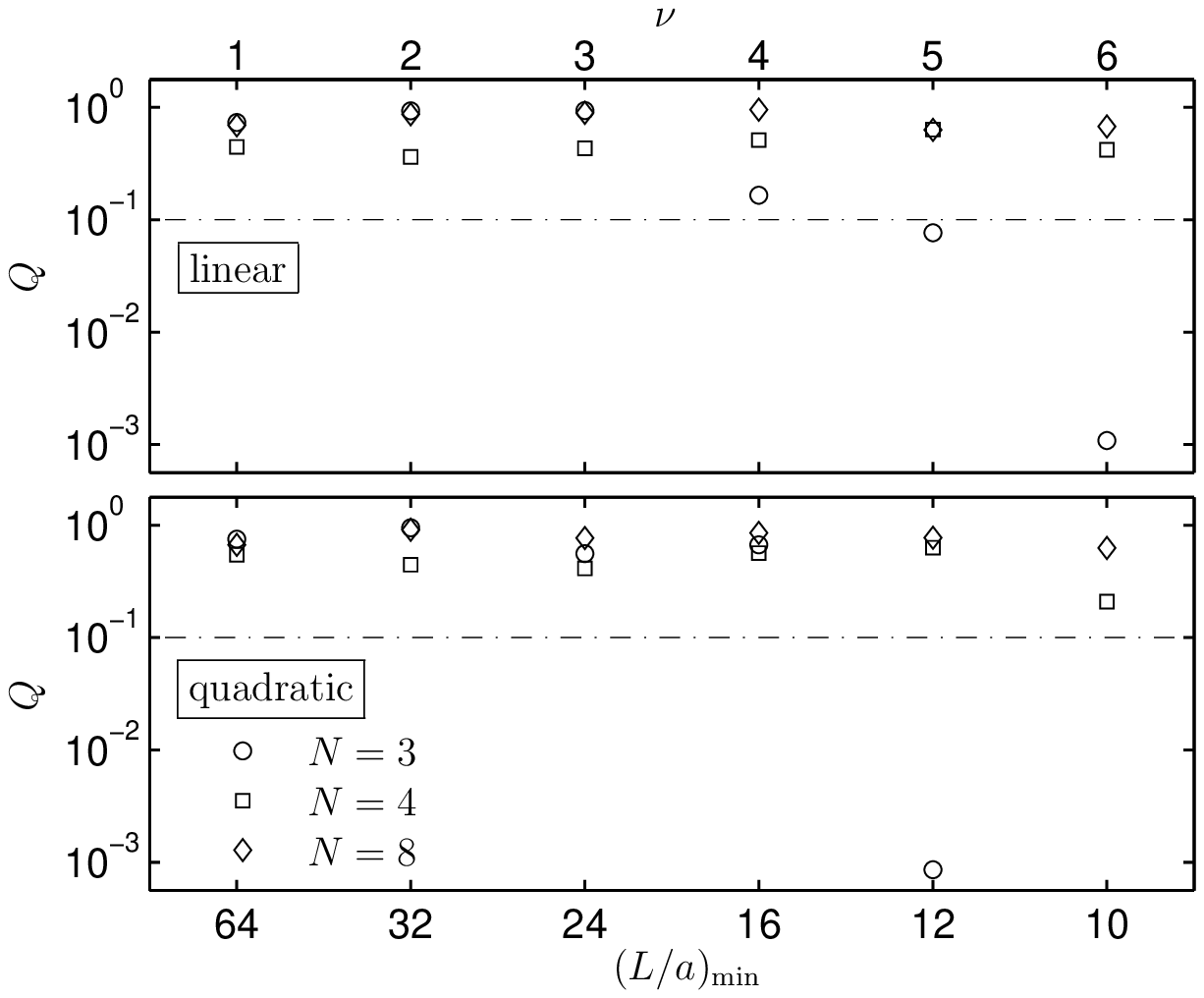}
	    \caption{$Q$-values of the fits to \eq{fitlog1} (linear) and \eq{fitlog2} (quadratic) 
	    against $(L/a)_{\mathrm{\min}}$. For $N=4\,,8$ both fits are perfect in the whole range. For $N=3$
	    both fits start to become unlikely at $(L/a)_{\mathrm{\min}}=12$.}
	    \label{fig_log}
    \end{figure}

In addition to fit A  with a general correction to scaling exponent we also tried
the Symanzik and the ad hoc linear form with both including the simplest logarithmic correction
\begin{eqnarray}
    \mathrm{B}: &  & c_0 + c_1\, (a/L)\phantom{^2} + c_2\, (a/L)\phantom{^2}\,\ln(a/L) \, .\label{fitlog1}\\
    \mathrm{C}: &  & c_0 + c_1\, (a/L)^2 + c_2\, (a/L)^2\,\ln(a/L) \, ,\label{fitlog2}
\end{eqnarray}
Note that at $N=\infty$ the form C is known to apply (see next section). 
In the upper and the lower subplot of \fig{fig_log} the $Q$-values are shown for the 
linear (\eq{fitlog1}) and quadratic (\eq{fitlog2}) form respectively. 
For $N=4,\,8$ both fits are tolerable in the whole $(L/a)_{\mathrm{\min}}$-range.
In the case $N=3$ and at small $(L/a)_{\mathrm{\min}}$ both forms become unlikely. 
For $(L/a)_{\mathrm{\min}}=16$  the form C, with quadratic lattice artifacts, gives the better fit, 
but for $(L/a)_{\mathrm{\min}}=12$ it is already intolerable whereas the 
linear fit with $Q\approx10^{-1}$ could perhaps still be accepted.

In summary, the single exponent ansatz A hints at `effective' values between one and
two in the range of cutoffs studied. With logarithmic modifications inluded, linear
and quadratic dependence is asymptotically about equally well compatible with our data.
All fits studied here have at most two free parameters. More complicated fits
can be extended to coarser lattices, and then the additional parameters are actually
largely determined by these additional included data. We do not think that much can be learnt
from such more involved fits. We rather turn now to the study of large $N$ to then apply the same
fitting procedure in a case where we 
have additional insight in the true cutoff dependence.

\section{O($N$) model at large $N$}

At finite $N$ Monte Carlo simulations were our tool to go beyond
perturbation theory.
In this section we complement this information by evaluating
the first two terms of the $1/N$ expansion of the step scaling
function. As we shall discuss they are {\em nonperturbative} in the
renormalized coupling $M(L)L$.
Related large $N$ computations are found for example in \cite{Caracciolo:1998gg} and 
\cite{Flyvbjerg:1990jg}, but to our knowledge the step scaling function as defined here
and in particular its subleading correction is not in the literature. 

For simplicity we employ lattice units in the following
with the lattice spacing set to $a = 1$. Scaling is then studied in $1/L$.

\subsection{Spin correlation function}

To extract the mass-gap we need to study the spin two-point function.
We Fourier represent the spin constraints in (\ref{Zsigma}) and add 
a source term to obtain
the generating functional of spin correlation functions
\begin{equation}
  Z ( J ) = \int \prod_x d^N s \frac{d \alpha}{2 \pi i} \mathe^{- S ( s ) - (
  \alpha, s^2 - 1 ) + ( J, s )}
\end{equation}
with each $\alpha ( x )$ integrated along the imaginary axis. We rescale
$\alpha$, shift contours $\alpha ( x ) = m_0^2 + i \varphi ( x )$ and perform
the Gaussian $s$-integrals to obtain
\begin{equation}
  Z ( J ) \propto \int \prod_x d \varphi \mathe^{- S_{\tmop{eff}} ( \varphi
  ) + \frac{\gamma}{2 N} ( J, K^{- 1} J )}
\end{equation}
with
\begin{equation}
  S_{\tmop{eff}} ( \varphi ) = \frac{N}{2} \tmop{tr} \log K - \frac{i N}{2
  \gamma} \sum_x \varphi
\end{equation}
involving the operator
\begin{equation}
  K = - \partial_{\mu} \partial_{\mu}^{\ast} + m_0^2 + i \varphi \equiv ( D^{-
  1} + i \varphi ) = D^{- 1} ( 1 + i D \varphi ),
\end{equation}
where the scalar field $\varphi$ is regarded as a diagonal matrix and
\begin{equation}
  \label{Dprop} D = ( - \partial_{\mu} \partial_{\mu}^{\ast} + m_0^2 )^{- 1} .
\end{equation}
The mass $m_{0^{}}^2$ is chosen such that $\alpha ( x ) = m_0^2$ is a
saddlepoint or equivalently, that $S_{\tmop{eff}} ( \varphi )$ is minimal
for $\varphi = 0$. This occurs if the gap equation
\begin{equation}
  \frac{1}{\gamma} = \frac{1}{V} \tmop{tr} K^{- 1} |_{\varphi = 0} =
  \frac{1}{V} \tmop{tr} D
\end{equation}
is fulfilled with the volume $V$ equaling the number of lattice sites.

Next we expand $S_{\tmop{eff}}$ in powers of $\varphi$, omit a constant and
rescale $\varphi \rightarrow \sqrt{2 / N} \varphi$ to find
\begin{equation}
  \label{Seff} S_{\tmop{eff}} ( \varphi ) = \left[ \frac{1}{2} \tmop{tr} ( D
  \varphi )^2 - \frac{i}{3} \sqrt{\frac{2}{N}} \tmop{tr} ( D \varphi )^3 -
  \frac{1}{2 N} \tmop{tr} ( D \varphi )^4 + \ldots . \right] .
\end{equation}
Similarly we then get
\[ K^{- 1} = D - i \sqrt{\frac{2}{N}} D \varphi D - \frac{2}{N} D \varphi D
   \varphi D + \ldots . \]
The spin two point function is given by
\[ \left\langle s_i ( x ) s_j ( y ) \right\rangle =
   \frac{\gamma}{N} \delta_{i j} G ( x, y ), \]
\[ G ( x, y ) = \left\langle K^{- 1} ( x, y ) \right\rangle \]
with the expectation value taken with $S_{\tmop{eff}}$ and employing an
obvious kernel notation for the operator $K^{- 1}$. Upon $1 / N$ expansion
this implies
\begin{equation}
  \label{G1} G = D - \frac{2}{N} \left\langle D \varphi D \varphi D
  \right\rangle_0 + \frac{2}{3 N}  \left\langle D \varphi D \tmop{tr} ( D
  \varphi )^3 \right\rangle_0 + O ( N^{- 2} )
\end{equation}
with $\left\langle \ldots \right\rangle_0$
taken now  with only the quadratic part of 
$S_{\tmop{eff}}$. We introduce
the selfenergy operator $H = \frac{1}{N} H_1 + O ( 1 / N^2 )$ by setting
\begin{equation}
  \label{self} G^{- 1} = D^{- 1} + H
\end{equation}
and read off by comparing with (\ref{G1})
\begin{equation}
  \label{self1} H_1 = 2 \left\langle \varphi D \varphi \right\rangle_0 -
  \frac{2}{3} \left\langle \varphi \tmop{tr} ( D \varphi )^3 \right\rangle_0
  \equiv H_{1, a} + H_{1, b} .
\end{equation}
Going to momentum space with
\begin{equation}
  \tilde{D} ( p ) = \frac{1}{\hat{p}^2 + m_0^2}, \quad \text{$\hat{p}_{\mu} =
  2 \sin ( p_{\mu} / 2 )$}
\end{equation}
and the $\varphi$-propagator $\tilde{W} ( p )$ determined in the first term in
(\ref{Seff}) as
\begin{equation}
  \tilde{W}^{- 1} ( p ) = \frac{1}{V} \sum_q  \tilde{D} ( p - q ) \tilde{D} (
  q ),
\end{equation}
the two contributions have the form
\begin{equation}
  \tilde{H}_{1, a} ( p ) = \frac{2}{V} \sum_q \tilde{D} ( q )
  \tilde{W} ( q - p )
\label{H1a}
\end{equation}
and
\begin{equation}
  \tilde{H}_{1, b} = - \tilde{W} ( 0 ) \frac{2}{V^2} \sum_{q,r} \tilde{W} ( q )
   \tilde{D} ( q-r ) \tilde{D}^2 ( r ) .
\label{H1b}
\end{equation}
    \begin{figure}[tb]
	    \centering
	    \includegraphics{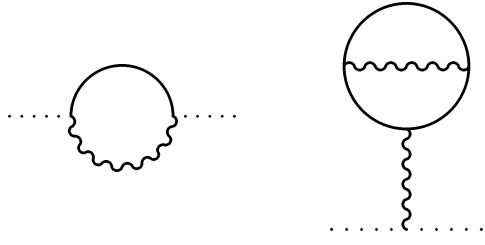}
	    \caption{Diagrammatic representation of (\ref{H1a}) and (\ref{H1b}).}
	    \label{Ndiagrams}
    \end{figure}
These terms are visualized by the diagrams in \fig{Ndiagrams} where $\tilde{D}$ is
represented by solid and $W$ by wiggly lines. Using the explicit form of
$\tilde{D}$ we can eliminate the double sum in the constant $\tilde{H}_{1, b}$ and
find
\begin{equation}
  \label{H1b1} 
  \tilde{H}_{1, b} = \tilde{W} ( 0 ) \frac{1}{V} \sum_q \tilde{W} ( q )
  \frac{\partial}{\partial m_0^2}  \tilde{W}^{- 1} ( q ) .
\end{equation}

\subsection{Mass gap}

For our two-dimensional volume $V = T L$ we now take the limit $T \rightarrow
\infty$. The momentum $p_0$ becomes continuous and we have to replace
\[ \frac{1}{V} \sum_p \ldots . \rightarrow \frac{1}{L} \sum_{p_1} \int \frac{d
   p_0}{2 \pi} \ldots . \]
The gap equation becomes in this limit
\begin{equation}
  \label{gap} \frac{1}{L} \sum_{p_1} \int \frac{d p_0}{2 \pi} 
  \frac{1}{\hat{p}^2 + m_0^2} = \frac{1}{L} \sum_{p_1} \frac{1}{2 \hat{\omega}
  ( p_1 ) \sqrt{1 + \hat{\omega}^2 ( p_1 ) / 4}} = \frac{1}{\gamma}
\end{equation}
with $\hat{\omega}^2 ( p_1 ) = ( \hat{p}_1^2 + m_0^2 )$. It furnishes a
one-to-one relation (at given $L$) between $m_0^2$ and $\gamma > 0$.

The mass gap $M > 0$ of the transfer matrix can be extracted from the two
point function in momentum space $\tilde{G} ( p )$ by finding the smallest
real $M$ with
\begin{equation}
  \label{pole} \tilde{G}^{- 1} ( p_M ) = 0, \quad p_M = ( i M, 0 ) .
\end{equation}
We expand
\begin{equation}
  \label{Mex} M = M_0 + \frac{1}{N} M_1 + \ldots .
\end{equation}
At leading order $G$ coincides with $D$ which yields
\begin{equation}
  \label{Mm} M_0 = \ln \left[ 1 + m_0 \sqrt{1 + m_0^2 / 4} + m_0^2 / 2 \right]
  = M_0 ( \gamma, L ) .
\end{equation}
If we now combine (\ref{pole}), (\ref{Mex}), (\ref{self}) we obtain for the
first correction to the mass
\begin{equation}
  M_1 = \frac{\tilde{H}_1 ( p_{M_0} )}{2 m_0 \sqrt{1 + m_0^2 / 4}} .
\end{equation}

In the practical evaluation $M_0 ( \gamma, L )$ follows easily from solving
(\ref{gap}) for instance by the Newton Raphson algorithm. For the correction
it is essential that also the $T \rightarrow \infty$ limit of $\tilde{W}$ can
be taken in closed form. In a first step we write it as a contour integral in
$z = \mathe^{i q_0}$ over the unit circle
\[ \tilde{W}^{- 1} ( p ) = \frac{1}{L} \sum_{q_1} \oint \frac{d z}{2 \pi i z} 
   \frac{1}{2 \cosh ( \theta ) - z - z^{- 1}}  \frac{1}{2 \cosh ( \delta ) - z
   z_0 - z^{- 1} z_0^{- 1}} . \]
Here we have introduced
\[ z_0 = \mathe^{- i p_0}, \quad \cosh ( \delta ) = 1 + \hat{\omega}^2 ( q_1 -
   p_1 ) / 2, \quad \cosh ( \theta ) = 1 + \hat{\omega}^2 ( q_1 ) / 2. \]
After some algebraic simplifications we arrive at
\begin{equation}
  \tilde{W}^{- 1} ( p ) = \frac{1}{2 L} \sum_{q_1} 
  \frac{\coth(\theta)+\coth(\delta)}{4\sinh^2 \left(\frac{\theta + \delta}{2} \right) +
  \hat{p_0}^2} .
\end{equation}
With this form the computation of both $\tilde{H}_{1, a} ( p_{M_0} )$ and
$\tilde{H}_{1, b}$ in (\ref{H1a}) and (\ref{H1b1})  leaves us with a single
numerical integration over $q_0 \in [ 0, \pi ]$ (using symmetry) in combination
with a two-fold summation over the spatial momenta. This is easily
feasible on a PC with high precision and up to $L$ of a few hundred.
Between the two diagrams a divergence $\propto L^2$ cancels\footnote{We thank Janos
Balog pointing this out.}.

A final remark concerns the continuation of $\tilde{H}_{1, a}$ to imaginary momentum.
For real momenta $p$ we can obviously use the shifted form
\begin{equation}
  \tilde{H}_{1, a} ( p ) = \frac{2}{L} \sum_{q_1} \int \frac{dq_0}{2\pi} 
  \tilde{D} ( q +r)
  \tilde{W} ( q - p +r) \nonumber
\end{equation}
instead of (\ref{H1a}) with any admissable lattice momentum $r$ (real $r_0$,
$r_1 =2\pi n/ L$) including
the case $r=p$. Only in the latter case we encounter a singularity if then we simply
substitute $p \to p_{M_0}$ and a more careful continuation would be required.
We used (\ref{H1a}) in its original form amounting to $r=0$.

\subsection{Step scaling function}

We can now determine the coefficients of the expansion (\ref{largeNSigma}).
For the leading term
we read off \cite{WeiszInt}
\begin{equation}
  \Sigma_0 ( u, L ) = 2 LM_0 ( \gamma, 2 L ) |_{LM_0 ( \gamma, L ) = u}
\end{equation}
with $\gamma$ determined by $u$.
For the next term we find (with the same $\gamma$)
\begin{equation}
  \Sigma_1 ( u, L ) = 2 L M_1 ( \gamma, 2 L ) - L M_1 ( \gamma, L )
  \frac{\partial \Sigma_0 ( u, L )}{\partial u},
\end{equation}
where the derivative is evaluated in the form
\begin{equation}
  \frac{\partial \Sigma_0 ( u, L )}{\partial u} = \frac{2 \frac{\partial M_0 (
  \gamma, 2 L )}{\partial \gamma}}{\frac{\partial M_0 ( \gamma, L )}{\partial
  \gamma}} .
\end{equation}
The derivatives $\partial M_0 / \partial \gamma$ follow from (\ref{Mm}) and
$\gamma$-derivatives on both sides of (\ref{gap}) avoiding the need to take
any derivative numerically.
 
Each term in the expansion
is expected to have a continuum limit for fixed $u$
\begin{equation}
  \sigma_i ( u ) = \lim_{L \rightarrow \infty} \Sigma_i ( u, L )
\end{equation}

\subsection{Infinite $N$}

\subsubsection{Continuum}

At leading order the step scaling function is given just by the gap equation.
In ref.\cite{Caracciolo:1998gg} we find asymptotic large $L$ expansions of lattice sums
relevant for (\ref{gap}) that yield
\begin{equation}
  \frac{2}{\gamma} = \frac{1}{\pi}  \left[ \ln L + G_0 ( u / 2 \pi ) \right] +
  \frac{1}{u} + C - \frac{u^2}{8 \pi^2} \frac{\ln L}{L^2} + \rO (1 / L^2 ) .
\end{equation}
Here $C$ is some constant and $G_0$ is given by
\begin{equation}
  \label{G0} G_0 ( \alpha ) = \sum_{n = 1}^{\infty} \left[ \frac{1}{\sqrt{n^2
  + \alpha^2}} - \frac{1}{n} \right] .
\end{equation}
Note that the Symanzik form of the leading lattice artifacts holds here for all
couplings. The exact continuum step scaling function $\sigma_0 ( u )$ is given
by the transcendental equation
\begin{equation}
  \frac{1}{\pi}  \left[ \ln 2 + G_0 ( \sigma_0 / 2 \pi ) \right] +
  \frac{1}{\sigma_0} = \frac{1}{\pi} G_0 ( u / 2 \pi ) + \frac{1}{u} .
\label{transcend}
\end{equation}
The function $G_0$ possesses a perturbative expansion, 
convergent\footnote{
This is closely related to the series for $M(L)/\Lambda_{\overline{\text{MS}}}$
discussed in \cite{Luscher:1982uv}.} for $\alpha
< 1$,
\begin{equation}
  \label{G0small} G_0 ( \alpha ) = \sum_{n = 1}^{\infty} \frac{( 2 n ) !}{( n!
  )^2} \zeta ( 2 n + 1 ) ( - \alpha^2 / 4 )^n
\end{equation}
which implies the expansion coefficients of $\sigma$ to all orders. The first
ones in 
\begin{equation}
\sigma_0 = u + s_0 u^2 + s_1 u^3 + \ldots 
\label{sigma0PT}
\end{equation}
are
\begin{equation}
  \label{sigcoeff} s_0 = \frac{\ln 2}{\pi}, \quad s_1 = s_0^2, \quad s_2 =
  s_0^3, \quad s_3 = s_0^4 - s_0 \frac{\zeta ( 3 )}{4 \pi^3}, \quad s_4 =
  s_0^5 - s_0^2  \frac{7 \zeta ( 3 )}{8 \pi^3} .
\end{equation}
In ref.\cite{Shin:1998bh} the four loop $\beta$-function at finite $N$ was computed for
the LWW coupling. Its $N \rightarrow \infty$ limit yields an expansion for the
step scaling function that coincides with the present series and from $s_3$ we
find for a constant that was determined numerically before the compatible
result $\chi_1 = - \zeta ( 3 ) = - 1.20 \ldots$. Note that for the
quantities considered here we confirm the interchangeability of the large $N$-
and the cutoff-limit leading to renormalized perturbation theory.

    \begin{figure}[tb]
	    \centering
	    \includegraphics[width=10cm]{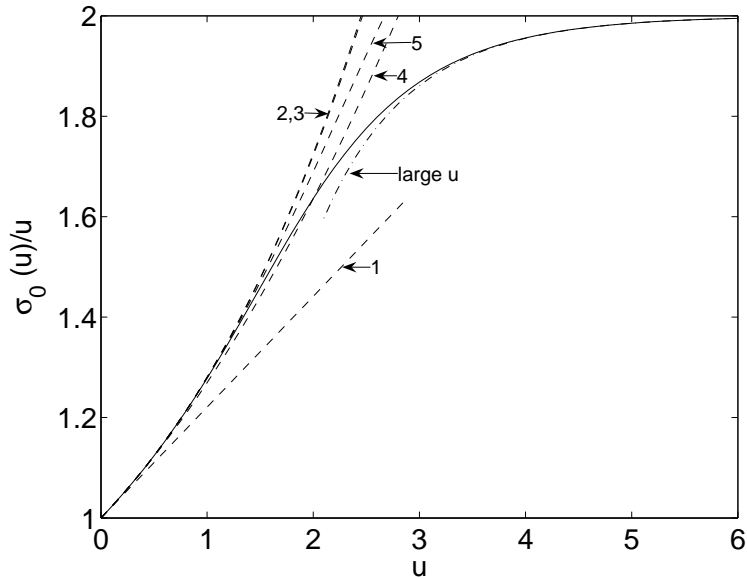}
	    \caption{Exact infinite $N$ step scaling function $\sigma_0$ together
            with perturbative and large $u$
            approximations. Labels $1,2,\ldots$ denote the expansion (\ref{sigma0PT})
            including  terms up to $s_0,s_1,\ldots$ and agree with the loop-order in conventional
            perturbation theory.}
	    \label{sigma0}
    \end{figure}
 
At large $\alpha$ the function (\ref{G0}) has an asymptotic expansion \cite{Caracciolo:1998gg}
\begin{equation}
  \label{G0large} G_0 ( \alpha ) = - \ln ( \alpha / 2 ) - \frac{1}{2 \alpha} +
  2 \sum_{n = 1}^{\infty} K_0 ( 2 \pi n \alpha ) + \gamma_{\rm E}
\end{equation}
with Euler's constant $\gamma_{\rm E}$ and the modified Bessel function
$K_0$. It implies the usual exponential corrections to the finite volume mass
gap
\begin{equation}
  \sigma_0 ( u ) \cong 2 u \left[ 1 - 2 K_0 ( u ) \right] \cong 2 u \left[ 1 -
  \sqrt{2 \pi / u} \exp ( - u ) \right] .
\end{equation}
In \fig{sigma0} we plot the exact continuum step scaling function together with its
small and large $u$ expansions.  We remark that a few orders of perturbation theory
approximate $\sigma_0$ well up to $u \lessapprox 1.5$.

Eq. (\ref{transcend}) can of course also be analyzed for other rescaling
factors then two. In particular from infinitesimal rescaling we obtain the
nonperturbative $\beta$-function in the form
\begin{equation}
\beta(u)=-\frac{u^2}{\pi}\frac{1}{1-\frac{u^2}{2\pi^2}G_0'(u/2\pi)}   +\rO(1/N).
\end{equation}

\subsubsection{Lattice artifacts}

At finite $L$ we compute $\Sigma_0 ( u, L )$ exactly by using (\ref{gap})
twice (with $L$ and $2 L$), eliminating $\gamma$ and expressing $m_0$ by $M_0$
via the inverse of (\ref{Mm}), $m_0 = 2 \, \sinh ( M_0 / 2 )$. This is
implemented numerically to machine precision. Another possibility is the small
volume expansion of the gap equation (\ref{gap}) yielding
\begin{equation}
  \label{gs} \frac{2}{\gamma} = \frac{1}{u}  \left( 1 + \sum_{n \geqslant 1}
  d_n ( L ) u^n \right) = \frac{1}{\Sigma_0}  \left( 1 + \sum_{n \geqslant 1}
  d_n ( 2 L ) \Sigma_0^n \right)
\end{equation}
with
\begin{equation}
  d_1 = \frac{1}{L}  \sum_{p_1 \neq 0} \frac{1}{\hat{p}_1 \sqrt{1 +
  \hat{p}_1^2 / 4}}, \quad d_3 = - \frac{1}{2 L^3}  \sum_{p_1 \neq 0} \frac{1
  + \hat{p}_1^2 / 2}{\hat{p}_1^3 ( 1 + \hat{p}_1^2 / 4 )^{3 / 2}},
\end{equation}
\begin{equation}
  \label{d24} d_2 = - \frac{1}{6 L^2}, \quad d_4 = \frac{7}{360 L^4} .
\end{equation}
Numerically we find by the method of App.~D of ref.\cite{Bode:2001jv}
\begin{equation}
  d_1 ( L ) = \frac{\ln L}{\pi} - 0.0703275870013 + 0.0872665 \frac{1}{L^2} +
  \rO( L^{- 4} )
\end{equation}
and
\begin{equation}
  d_3 ( L ) = - 0.0048460224504 - \frac{1}{8 \pi}  \frac{\ln L}{L^2} +
  0.0220539 \frac{1}{L^2} + \rO( L^{- 4} )
\end{equation}
with errors expected to lie beyond the quoted digits.

    \begin{figure}[tb]
	    \centering
	    \includegraphics[width=10cm]{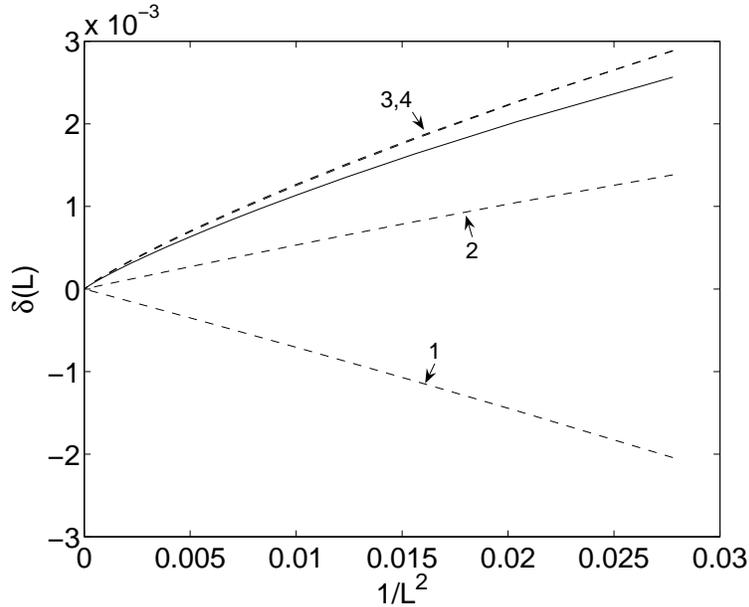}
	    \caption{Deviation of $\Sigma_0$ from the continuum limit at $u_0=1.0595$.}
	    \label{deltaLa}
    \end{figure}
    \begin{figure}[tb]
	    \centering
	    \includegraphics[width=10cm]{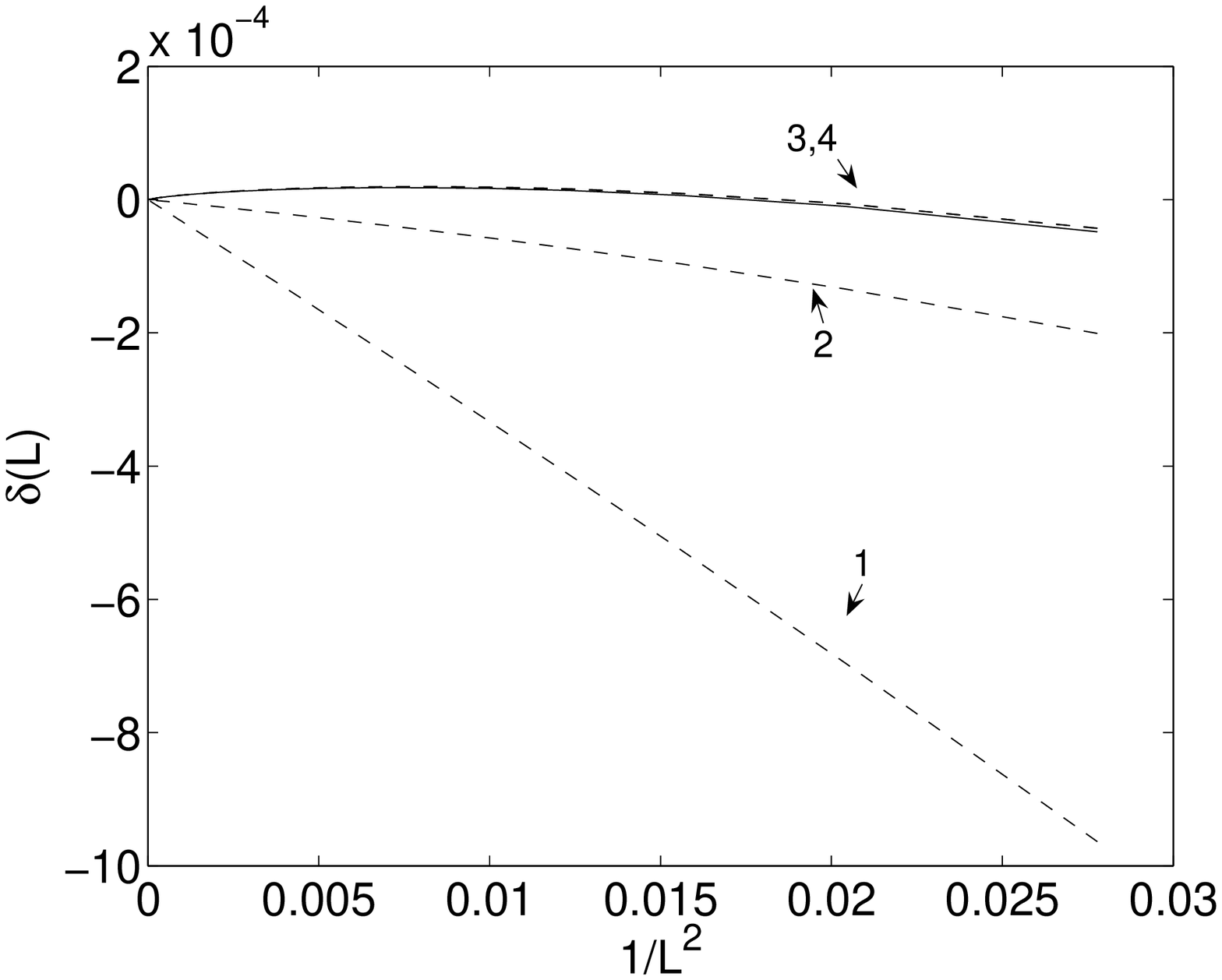}
	    \caption{As \fig{deltaLa}, but for $u=0.5$.}
	    \label{deltaLb}
    \end{figure}

In view of applications of the step scaling technique for QCD it is
interesting to compare the exact lattice artifacts with their perturbative
expansion. We form the relative deviation
\begin{equation}
  \delta ( u, L ) = \frac{\Sigma_0 - \sigma_0}{\sigma_0} = \delta_1 ( L ) u +
  \delta_2 ( L ) u^2 + \ldots .
\label{deltadef}
\end{equation}
Setting
\begin{equation}
  \Delta d_i ( L ) = d_i ( 2 L ) - d_i ( L )
\end{equation}
and combining series we find
\begin{equation}
  \delta_1 ( L ) = \Delta d_1 ( L ) - s_0,
\end{equation}
\begin{equation}
  \delta_2 ( L ) = \Delta d_2 ( L ) - s_1 + s_0^2 + \Delta d_1 ( L ) \delta_1
  ( L ),
\end{equation}
and similar but lengthy expressions for $\delta_{3, 4} ( L ) .$ In \fig{deltaLa} the
exact $\delta ( L )$ is plotted together with perturbative approximations at
$u_0 = 1.0595$. While the continuum limit in this range is well described by
perturbation theory, this is not really the case for $\delta$. The same was
observed in \cite{stepscaling} at $N = 3$. 
The leading term does not even have the right
sign. This situation improves at yet smaller couplings as shown in \fig{deltaLb}.
Heuristically, with $\delta$ being a short distance quantity, we thought it
might be a good idea to reexpand it in the bare coupling $\gamma ( u, L )$
which is always known nonperturbatively in computations of step scaling
functions. This experiment did however not lead to a more accurate
perturbative description of $\delta$ in the present case.

\subsection{$1 / N$ correction}

For a number of $u$-values we numerically
determined $\Sigma_1 ( u, L )$ up to $L = 128$ and higher for $u_0$.
Larger values are possible with some effort, but are of limited use unless
also the precision is raised beyond 64 bits. 
Sofar we have not succeeded in analytically extracting the large $L$-behavior
and therefore perform a numerical analysis here.

The expected form is
\begin{equation}
  \Sigma_1 ( u, L ) = \sigma_1 ( u ) + \frac{1}{L^2} \Sigma_1^{(2)} ( u,L)
  + \frac{1}{L^4} \Sigma_1^{(4)} ( u,L) + \rO(L^{-6})
\end{equation}
where the $\Sigma_1^{(k)}$ have an only weak residual $L$-dependence. For the entirely
analogous expansion of the leading $\Sigma_0$ it rigorously follows from 
\cite{Caracciolo:1998gg} that  $\Sigma_0^{(2)}$ is a linear and $\Sigma_0^{(4)}$ a quadratic
polynomial in $\ln(L)$.

In terms of the symmetric difference
\begin{equation}
\tilde{\Delta} f(L)= \frac12 (f(L+1)-f(L-1))
\end{equation}
we define similarly to \cite{Luscher:1985wf} `blocked functions'
\begin{eqnarray}
B_i^{(2)} &=& -\frac12 L^3 \tilde{\Delta} \Sigma_i\\
B_i^{(4)} &=& -\frac1{16} L^4 (2+L\tilde{\Delta})^2 L\tilde{\Delta} \Sigma_i
\end{eqnarray}
for $i=0,1$. The reasoning behind this construction is that in $B_0^{(2)}$ the
continuum part of $\Sigma_0$ has been canceled and the $L^{-2}$ components
are promoted to order unity. The normalization is such that the coefficient of the term linear in $\ln L$
in $\Sigma_0^{(2)}$ is the same as in $B_0^{(2)}$. Similarly in
$B_0^{(4)}$ the $L^{-2}$ components are canceled and the 
coefficient of $(\ln L)^2/L^4$ is preserved.
Although the a priori form of $\Sigma_1^{(k)}$ is not analytically known sofar, 
we construct
$B_1^{(k)}$ in the same way.
    \begin{figure}[tb]
	    \centering
	    \includegraphics[width=10cm]{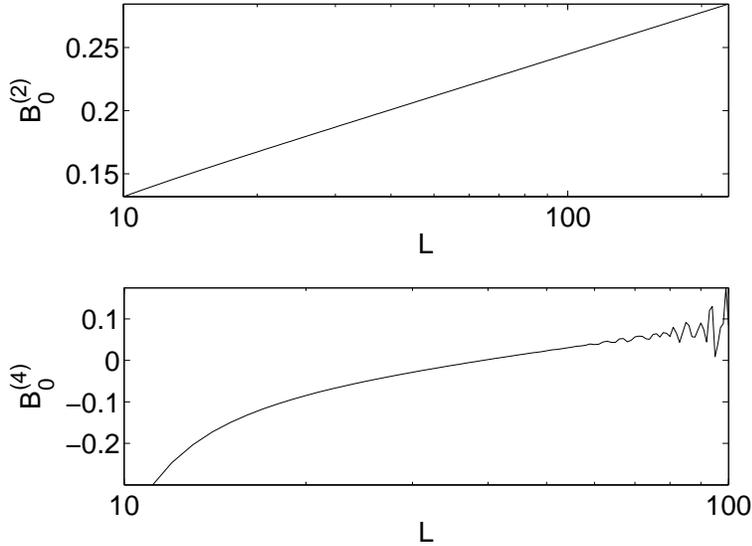}
	    \caption{Variation of $L^{-2}$ and $L^{-4}$ 
              components in $\Sigma_0(u_0,L)$.}
	    \label{asym0}
    \end{figure}
    \begin{figure}[tb]
	    \centering
	    \includegraphics[width=10cm]{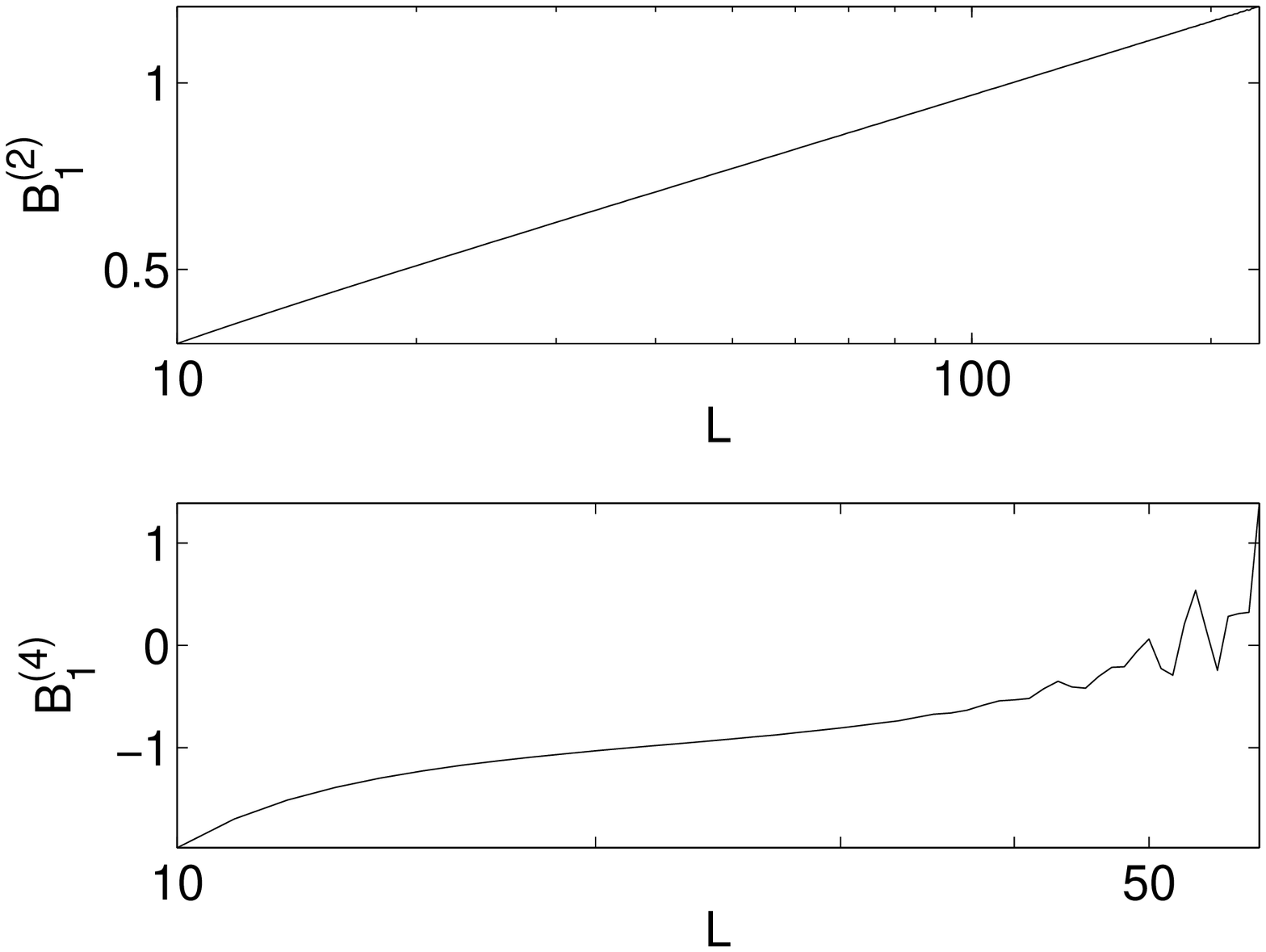}
	    \caption{As \fig{asym0} but for the subleading term $\Sigma_1$}
	    \label{asym1}
    \end{figure}
In \fig{asym0} we plot $B_0^{(k)}$ at $u=u_0$ for a range of $L$ and see the
expected slowly varying $L$ dependence. The analogous plot \fig{asym1} for
the $1/N$ correction shows a rather similar structure.
Both lower plots eventually start to become irregular at large $L$ due to roundoff
noise. For both terms $\Sigma_i^{(2)}$ looks strictly linear in $\ln L$ and
the $(\ln L)^2$ components in $\Sigma_i^{(4)}$, i.~e. the curvature at large $L$,
seems to be very weak. For other $u$-values the situation is qualitatively similar,
although coefficients are larger for larger $u$.

\begin{figure}[tb]
	    \centering
	    \includegraphics[width=10cm]{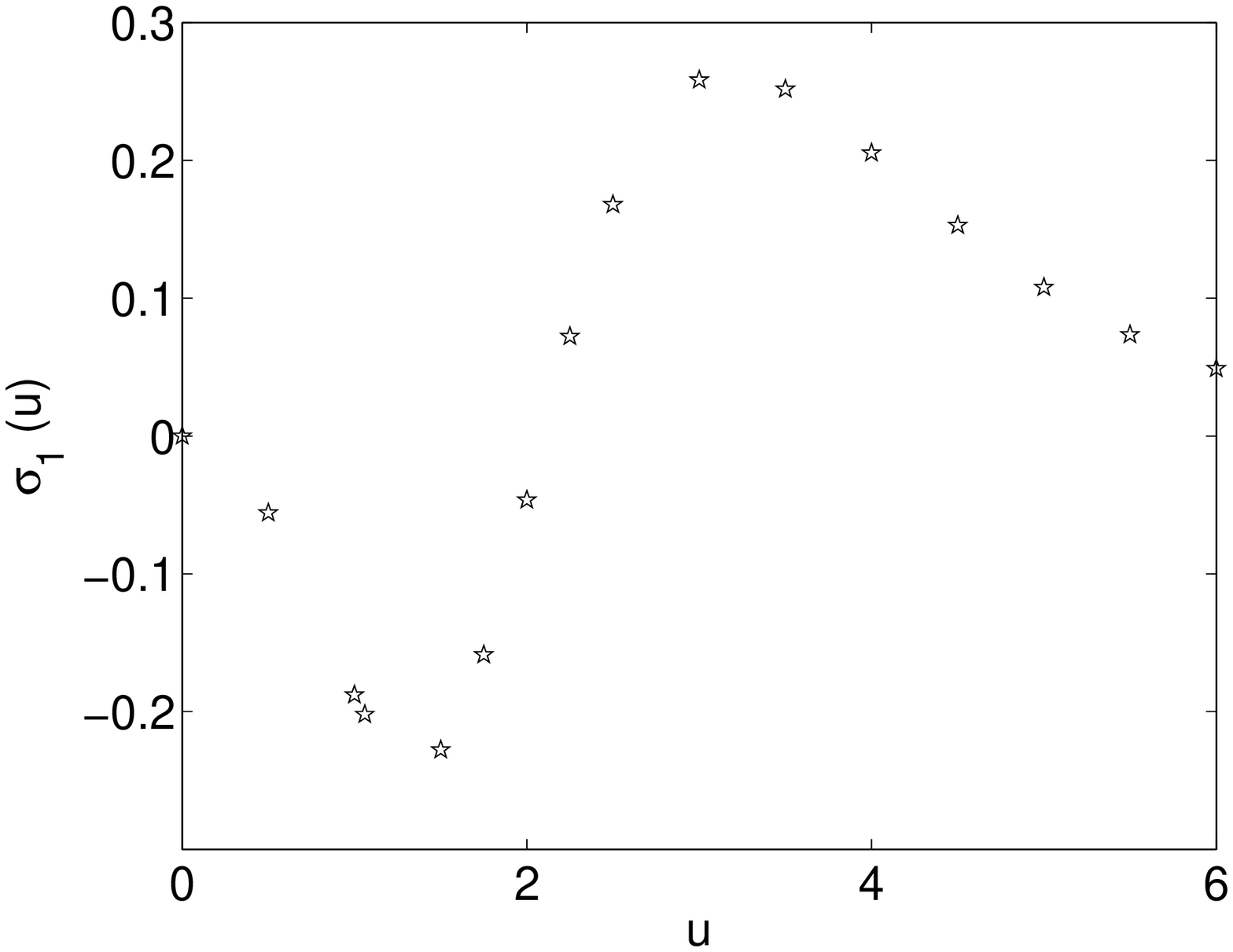}
	    \caption{Leading correction $\sigma_1(u)$.}
	    \label{sig1plot}
    \end{figure}

After this more qualitative investigation we
obtain the coefficients in the asymptotic expansion
\begin{equation}
\Sigma_1(u,L)=\sigma_1(u)+A(u)\frac{\ln{L}}{L^2}+B(u)\frac{1}{L^2}+\rO(L^{-4})
\label{Sig1fit}
\end{equation}
by the method 
introduced in \cite{Bode:2001jv}.
The resulting
numbers are collected in Tab.\ref{sigma1} and the continuum values
are plotted in \fig{sig1plot}. The correction obviously has to (and does) vanish
for $u\to 0,\infty$ since $\sigma(u,N)$ assumes $N$-independent trivial
limits $u$ and $2u$.

It seems rather difficult but perhaps not hopeless to analytically derive
the asymptotic $L$-dependence of $\Sigma_1$. We hope to come back to such an analysis
in a future publication \cite{JBUW}. 

\begin{table}[htb]
	\centering
	\begin{tabular}{llll}
	$u$	& $\sigma_1$		& $A$		& $B$  \\ \hline
  0.5 &           -0.055886    & 0.012 & -0.066\\
    1 &           -0.187974    & 0.220 & -0.184\\
1.0595&           -0.202161    & 0.284 & -0.200 \\
  1.5 &           -0.227820    & 1.27  & -0.442\\
 1.75 &           -0.158761    & 2.23  & -0.84\\
    2 &           -0.046463    & 3.26  & -1.56\\
 2.25 & \phantom{-}0.072185    & 4.13  & -2.52\\
  2.5 & \phantom{-}0.167824    & 4.67  & -3.53(1)\\
    3 & \phantom{-}0.258377    & 4.82  & -5.12(1)\\
  3.5 & \phantom{-}0.251671(1) & 4.21(1) & -5.71(3)\\
    4 & \phantom{-}0.205320    & 3.38(1) & -5.50(4)\\
  4.5 & \phantom{-}0.152853    & 2.58(1) & -4.85(4)\\
    5 & \phantom{-}0.107791    & 1.92(1) & -4.04(4)\\
  5.5 & \phantom{-}0.073422    & 1.40(1) & -3.24(4)\\
    6 & \phantom{-}0.048858    & 1.01(1) & -2.52(4)\\
	\end{tabular}                                          
	\caption
		{Continuum extrapolation of $\Sigma_1$, see eq.(\ref{Sig1fit})}
	\label{sigma1}
\end{table}
 
\begin{figure}[tb]
	    \centering
	    \includegraphics[width=10cm]{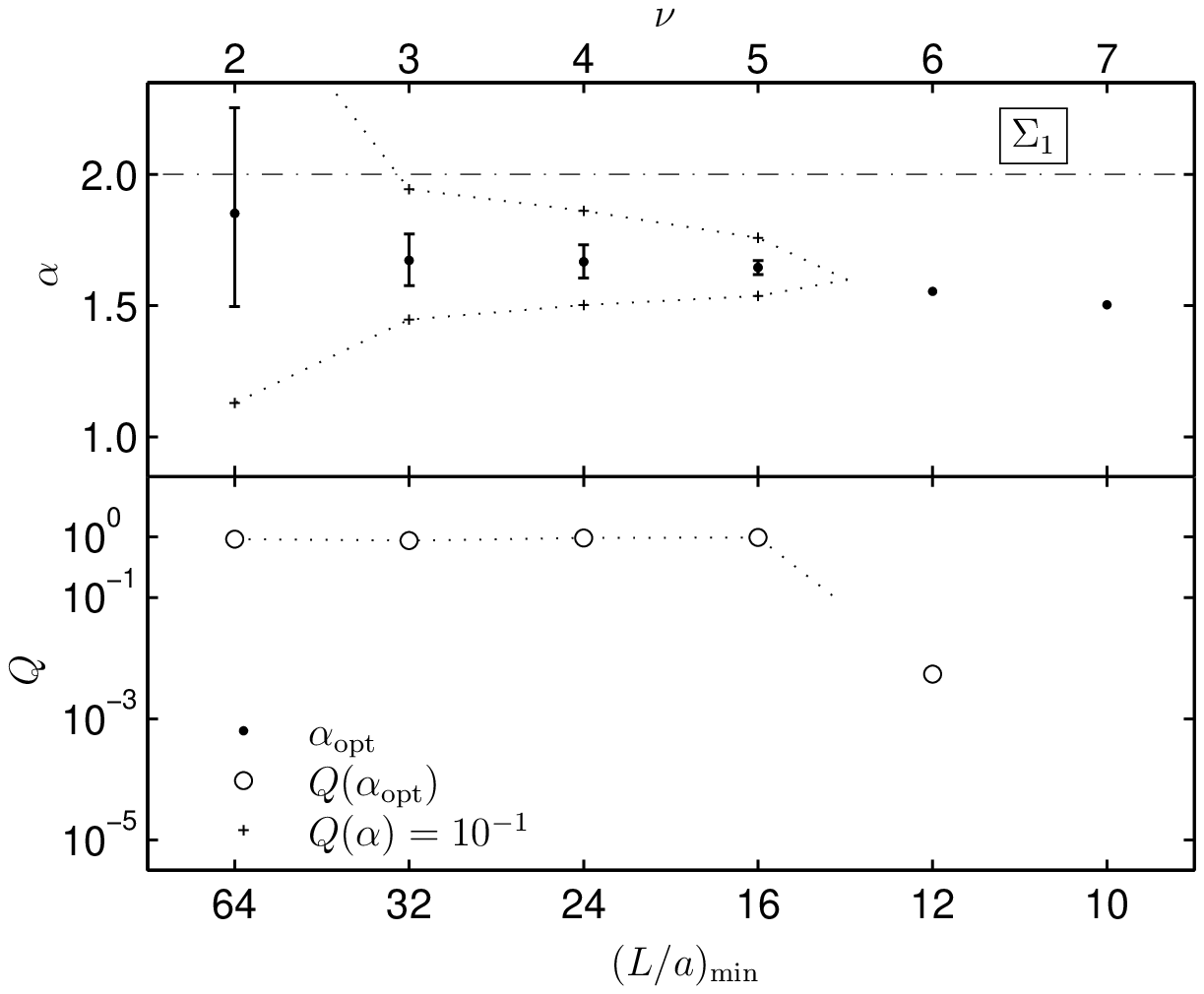}
	    \caption{Fit A for the large $N$ data $\Sigma_1(u_0,L)$ with artificial errors added.}
	    \label{fuzz}
    \end{figure}

It is instructive to perform with our large $N$ data the same analysis
as for the Monte Carlo data. For this purpose we `fuzz' our data with
artificial errors. For instance we add to $\Sigma_1(u_0,L)$
in the range $10 \le L \le 128$
Gaussian random numbers of width $2 \times 10^{-5}$.
This size is chosen such that the lattice artefact $|\Sigma_1(u_0,10)-\sigma_1(u_0)|$
of the $1/N$ correction gets a relative `statistical' error similar to the $N=3$ case.
The resulting plot analogous to \fig{fig_ssf3_aopt_imp} is shown in \fig{fuzz}.
We see a rather similar picture for this case where we are confident of the
lattice artefacts really having the Symanzik form. Very similar plots can also
be generated from the leading $N=\infty$ term. Also in \fig{fig_log} the new data points
would resemble the structure exhibited by the Monte Carlo data.
\section{Conclusions}\label{concl}

In this work we have further investigated the scaling violations of the
finite volume mass gap step scaling function of the O($N$) model at various $N$.
The discretizations are restricted to the standard action. The behaviour at
$N=3$ is confirmed by our added data and by our method of analysis.
With fits of the form (\ref{fitaopt}) a rather large number of coarser lattices
has to be excluded, and then a cutoff dependence with an exponent $1< \alpha<2 $
seems to be suggested in the accessible range of lattice spacings.
With one additional logarithmic correction both exponents 1 (ad hoc) and 2 (Symanzik) 
(\ref{fitlog1},\ref{fitlog2})
can be used
with again only the finer lattices included. For $N=4,8$ a similar but less
pronounced picture emerges. Our sensitivity is weaker
since the lattice artefacts are generally much smaller and hence harder to measure.
Although there may be a trend of `normalization' with larger $N$, also at 
$N=4$ our data point toward $\alpha<2$ unless all lattices with $L/a<24$ are excluded.

It would be very interesting if theoretical principles could be found to organize
a leading log summation of leading cutoff effects in analogy to the renormalization
group improvement of continuum results. Such an attempt is here left to future study.

At $N=\infty$ it is rigorously known due to \cite{Caracciolo:1998gg} that cutoff effects are
of the Symanzik type with just $1/L^2, \ln(L)/L^2$ contributions. The leading
$1/N$ correction to the step scaling function, worked out here for the first time,
somewhat to our surprise
seems to have just the same behaviour, 
although this can be only demonstrated numerically at present.
Needless to say, we cannot exclude other admixtures if they are only
small enough. When these exact results are analyzed in the same way as
the Monte Carlo data, a qualitatively similar picture emerges.
It is hence very clear that we cannot conclude a truly anomalous
scaling behavior from the latter.

The large $N$ sigma model proved to be a very interesting case, 
also because we here have
an analytic handle on quantities that are nonperturbative in the coupling
of an asymptotically free theory. This is true for both the continuum and at finite
lattice spacing. This simplified laboratory can certainly be useful to study other issues.

\noindent {\bf Acknowledgements.}  We would like to thank Janos Balog, Burkhard Bunk, 
Tomasz Korzec and Peter Weisz
for helpful discussions. This work was supported by the Deutsche Forschungsgemeinschaft
in the form of Graduiertenkolleg GK~271 and Sonderforschungsbereich SFB~TR~09.

\clearpage

\bibliography{references}
\bibliographystyle{h-elsevier}

\end{document}